# A Design Science Method for Emerging Decision Support Environments

**Nguyen Hoang Thuan**
School of Information Management
Victoria University of Wellington
Wellington, New Zealand
Thuan.Nguyen@vuw.ac.nz

**Pedro Antunes**
School of Information Management
Victoria University of Wellington
Wellington, New Zealand
Pedro.Antunes@vuw.ac.nz

**David Johnstone**
School of Information Management
Victoria University of Wellington
Wellington, New Zealand
David.Johnstone@vuw.ac.nz

## Abstract

Emerging technologies and business models require organisations to continuously deal with complex, dynamic and unstructured issues, leading to the need for newer forms of decision support systems (DSS). However, in emerging environments the existing knowledge base can be scattered, unstructured, and sometimes conflicting, which challenges any efforts in designing DSS. This paper highlights the role of design science methods in developing these emerging areas, and suggests a design science method that focuses on consolidating the knowledge base by ontologically grounding experience and expertise. The proposed method is illustrated in the context of a published case, and validated by practically applying it to develop a crowdsourcing decision tool. The study contributes with recommendations on how to consolidate the knowledge base and design DSS artefacts in areas lacking strong theoretical foundations, and where expertise and experience are dominant sources of knowledge.

## 1 Introduction

Decision Support Systems (DSS), despite of its long history in the Information Systems (IS) discipline, is still an interesting 'alive and well' research area. This is because new technologies and business models are continuously emerging, which involve new business forms, large amounts of information, complex and unstructured issues. Thus, newer forms of decision-support development are continuously demanded (Hosack et al. 2012). This demand can be seen via many calls for further DSS in the emerging areas, like big data, social media, mobile computing, and crowdsourcing (Arnott and Pervan 2014; Geiger and Schader 2014), which provide great opportunities for DSS research. On the other hand, they raise challenges on how to rigorously develop DSS when promoting unestablished business structures and may lack strong theoretical foundation.

Aligning with the high percentage of DSS research adopting the design science paradigm (Arnott and Pervan 2014), we suggest that design science should play important role in developing DSS in the emerging environments. There are (at least) three reasons for this suggestion. First, design science emphasises a rigorous approach to advance current knowledge on design and development (Hevner et al. 2004). This is highly applicable to emerging DSS environments, where decision-support tasks involve consolidating domain knowledge for better decision (Nemati et al. 2002). Second, design science aims at developing innovative artefacts to address unstructured issues, which are also the major target of DSS research in the emerging areas. Third, design science is more focussed on utility than truth (Hevner and Chatterjee 2010), aligning to the purpose of support and improvement of DSS (Arnott and Pervan 2012). All in all, this combination of rigor, innovation and utility places design science as an appropriate paradigm to guide research addressing the emerging DSS environments.

Several methods for guiding design science research have already been proposed in the IS discipline, covering from broad principles and guidelines (Hevner et al. 2004) down to prescriptive accounts on how to plan (Peffers et al. 2007) and manage a design project (Hevner and Chatterjee 2010). Nevertheless, the application of several existing design methods may not to be suitable in emerging DSS environments, where the large wickedness and dynamism exist. Many existing methods require either established theories or meta-artefacts related to the problem (e.g. Carlsson et al. 2011; Pries-Heje and Baskerville 2008). However, this requirement is hardly met in emerging areas, as explained by Paré et al. (2015) that in such emerging issues/areas "an accumulated body of research exists but there is a lack of appropriate theories or current theories are inadequate in addressing existing research problems" (p. 188). Furthermore, the existing methods do not address the issue of lacking common understanding due to the emerging unestablished nature of the areas. For instance, when designing decision tools for crowdsourcing, one needs to address the inconsistent definitions existing in the field (Thuan et al. 2014). Without a common understanding about related concepts, there is a high risk of providing irrelevant and/or incomplete supports for the decision makers. As a result, in emergent environments we aligned with Miah et al. (2014) saying that "traditional DSS development methods have several limitations in supporting businesses, including conceptual mismatches, static models and inflexibility" (p. 273).

Given this situation, we highlight the importance of developing design science methods specifically tailored to emerging DSS environments, where the knowledge base is characterised by scattered knowledge sources rather than accepted theory. A similar suggestion has been made by Hevner and Chatterjee (2010), who note that "to insist that all design decisions and design processes be based on grounded behavioural or mathematical theories may not be appropriate or even feasible for a truly cutting-edge design artefact" (p. 18). More specifically, Hevner and Chatterjee (2010) suggest that design science research can be rigorously founded on three types of knowledge sources: 1) scientific theories and methods; 2) experience and expertise; and 3) meta-artefacts. In the emerging and dynamic DSS environments, we suggest a design method addressing the second type of knowledge source, i.e., building an ontological knowledge base from existing experience and expertise.

Our research proposes a design method guiding DSS artefact development in emerging environments. This method is based on experience and expertise reported in both scientific and practical publications to provide an ontological knowledge base for artefact development. In particular, the method applies an evidence-based strategy to scope and understand the current state-of-the-art on a particular topic (Paré et al. 2015), and uses conceptual models to explore and frame the DSS issue (Webster and Watson 2002). To consolidate the domain knowledge, the method suggests building ontologies to provide a framework for structuring the knowledge base (Corcho et al. 2003). Based on the constructed knowledge base, the artefact supporting decision makers in the DSS issues is promoted. The proposed method is named SCOA (**S**coping knowledge source, **C**onceptualisation, **O**ntology, and DSS **A**rtefact).

This research contributes to current knowledge by identifying the needs for design methods in immature emerging areas where a knowledge base cannot be based on existing theory but instead on accounts of experience and expertise. Another contribution includes a method and a set of guidelines for DSS research targeting these areas. Following Peffers et al. (2007), detailed steps of the method are illustrated and validated by applying retroactively to a published study and proactively to a crowdsourcing research project. This application serves to demonstrate the method's utility. A key distinction of the proposed method is its ability to consolidate scattered evidence into ontological structures, and thus enables the building of the knowledge base necessary for designing DSS artefacts in the emerging areas (Miah et al. 2014).

## 2   The Need of Design Methods in Emerging Environments

The IS field has many promising and emergent areas where knowledge should be advanced in spite of the lack of theoretical scaffolding (Hevner and Chatterjee 2010; Paré et al. 2015). As the role of DSS in emergent areas was already discussed in the introduction section, this section focuses on the needs of design methods in these areas, starting by analysing the characteristics of IS emerging fields. After reviewing two emergent fields, crowdsourcing and business process modelling (BPM) (Jonnavithula et al. 2015; Thuan et al. 2014), we agree with Hevner and Chatterjee (2010) that the knowledge in these areas may exist as experience and expertise. More precisely, our reviews could not find prevailing theories, something that was also noted by other authors (Zhao and Zhu 2014). Rather than theory, the reviews could find in the extant literature a relatively large amount of experience and expertise knowledge. For instance, the review on BPM literature found that a large portion of reviewed sources reported declarative statements, expert opinion, and technology design experience.

Furthermore, we stress two other important characteristics of these fields. First, these fields are characterised by various types of relevant contributions reported in the form of case studies, design studies, usability studies, and other engineering contributions. This is a logical situation since low-theory domains are often fruitful areas for technology development applications (Gregor and Jones 2007). Second, because of the theoretical immaturity of these domains, diverse and sometimes conflicted views and conceptualisations can be found in the knowledge base. For instance, in the field of crowdsourcing, we have found conflicted research findings regarding factors affecting the motivation to participate in crowdsourcing tasks (e.g. Brabham 2010; Kaufmann et al. 2011). As a result, the application of any research method for exploring these areas should take into account these distinctive characteristics.

Research in the emerging areas can adopt either behavioural science or design science, which are two major research paradigms in IS (Hevner and Chatterjee 2010). While behaviour science promotes several research methods for exploratory research, including grounded theory, phenomenography, and case study, it is not rare that design science can also be adopted. This is because design science targets wicked problems that normally emerge in such ill-defined environmental contexts (Pries-Heje and Baskerville 2008). Furthermore, many emerging technologies like crowdsourcing and social network spread from practical to academic worlds. This emphasises the choice of design science, which has strong links to practice, as stated by Hevner and Chatterjee (2010) that "design science research often begins by identifying and representing opportunities and problems in an actual application environment" (p. 17). Another advantage of applying design science in these areas is its varied forms of possible contributions, including constructs, models, methods, and instantiations (Gregor and Hevner 2013; Hevner et al. 2004).

In spite of the high applicability of design science to emerging areas, finding appropriate design methods that can provide methodical and transparent accounts of researchers' activities is challenging. To clarify this challenge, we adapted a framework proposed by Gregor and Hevner (2013), which represents possible applications and contributions of design science, in order to analyse the appropriateness of design methods. This framework is comprised of four quadrants, allocating within two dimensions: (problem) domain maturity and solution maturity. Three quadrants (2 to 4) characterised by low domain maturity and/or low solution maturity are areas where design science possibly maximises its contributions. The only quadrant that rarely requires design science is routine design where existing knowledge for both the problem and the solution is well-established (Gregor and Hevner 2013).

While agreeing with the framework, we added to it the characteristics of emergent environments (Figure 1). As already mentioned, these environments are characterised by availability of experiences and expertise but lacking of established theories. Thus, the x-axis was adapted to distinguish domain maturity into two categories: well-established and emergent domains, corresponding to availability of scientific theories, and experiences and expertise respectively. The y-axis is kept as the original solution maturity "that exist as potential starting points for solutions to the research question" (Gregor and Hevner 2013, p. 345). As these solutions may have different natures in each of the four quadrants, we named these solutions as normative, descriptive, prescriptive, and innovative. As a result, the adapted framework allows us to classify elements in the knowledge base and position the possible uses of the related design science methods. The framework is presented in Figure 1.

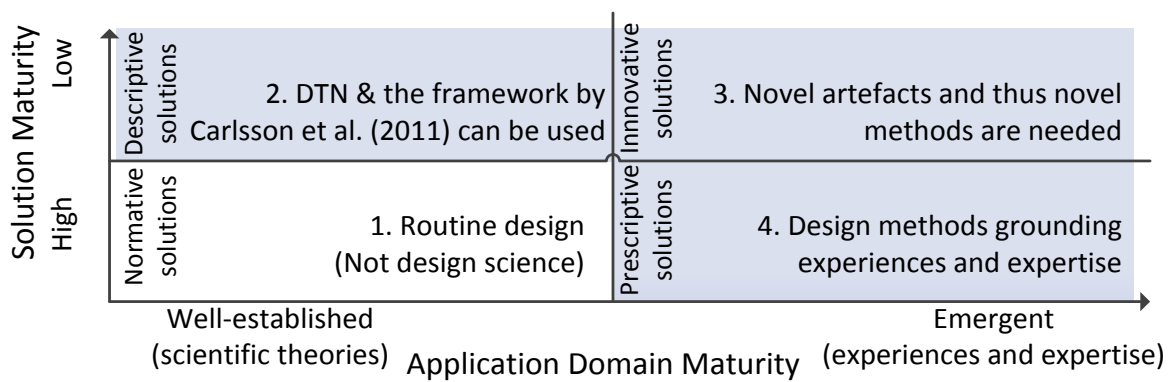

Figure 1: Possible uses of design science methods (adapted from Gregor and Hevner 2013)

In Figure 1, quadrant 1 considers well-established domains, with the mature theories and available solutions that are normatively applicable to the design problem. This quadrant is inadequate for design science research due to the lack of both challenging problems and the need for innovative solutions (Gregor and Hevner 2013). In quadrant 2, design science is applied in a mature domain for which established theories are expected to exist. However, these theories may not be directly relevant to the design problem, and thus may need to be analytically adapted. We note that if such theories can be directly applied to the design problem, quadrant 2 would become routine design. Given pre-existent theories, solutions in this quadrant are mainly descriptive according to existent theories, and theory-centred methods like the ones developed by Carlsson et al. (2011) and Design Theory Nexus (DTN) (Pries-Heje and Baskerville 2008) should be adopted. Quadrant 3 considers first-of-a-kind artefacts providing radical breakthroughs in the field, which usually demands open-ended and novel methods.

Quadrant 4 concerns the emerging research environment. In quadrant 4, many practical and engineering solutions may have already been applied and validated, but the field has not yet matured, which may have precluded theory building. We note that validated solutions here are not the solutions for the current design problem, but research outcomes and applications for other related problems (Gregor and Hevner, 2013). Given that, experiences gained from these applications and expert opinions on how to analyse, adapt, and solve the related problems are expected to be available. Thus, following Gregor and Hevner (2013), we suggest that in this quadrant, the knowledge bases may exist as individual experience, though such sources of knowledge may be scattered and lacking coherence and conceptual strength. As a result, design methods should not be theory-centred. Rather, existing experience and expertise can be used as sources of knowledge (Hevner and Chatterjee 2010). Following this assertion, we therefore highlight a need for design methods in these emerging areas.

However, the current literature shows no specific methods guiding design science research in this quadrant. Addressing this gap, the current study aims at proposing a design framework that can be applied to the immature, emergent DSS domains. Given the availability of experience and expert opinions in these domains, the proposed method grounds these knowledge sources to form a solid knowledge base of the design problem. According to Hevner and Chatterjee (2010), a design science knowledge base can be drawn from three sources: scientific theories and engineering methods, existing artefacts and processes, and experiences and expertise that "define the state of the art in the application domain of the research" (Hevner and Chatterjee 2010, p. 18). Although many theory-centred methods exist, our work is the first attempt to base a design method on articulated knowledge comprising experiences and expertise.

## 3   Conceptual Approach

The current study adopts the design science paradigm to develop our method. As already discussed in the background section, design science has several methods that can provide a rich source of knowledge on how to develop this artefact. In particular, Peffers et al. (2007) offer useful insights about the rationale and the conceptual steps necessary to build the artefact. Thus, we will closely follow Peffers et al.'s suggested six steps: 1) problem definition; 2) development objectives; 3) artefact development; 4) demonstration; 5) evaluation and 6) communication. The first three steps are discussed in this section, and the results – the SCOA method – will be presented in section 4. After that, demonstration and evaluation are presented. The final communication step is carried out through the current manuscript.

**Problem Definition.** The lack of design methods adequate for developing DSS artefacts in emergent areas was identified and discussed in the introduction and background sections. This is the main problem driving the method development. From that discussion, we further state that the method should guide a research project through the process of gathering and structuring expertise and experience knowledge on a particular DSS problem.

**Development Objectives.** The method development should be guided by the following three objectives. First, it should support artefact development in low-maturity DSS domains, and allow researchers to utilise experience and expertise sources of knowledge, including empirical studies, case studies, proof-of-concept developments, and best practices. This forms a knowledge base that enables the necessary reasoning (i.e. factors, concepts, relationships, and business rules) to build the knowledge component of the corresponding DSS. Second, the method should resolve the conflicting views and lack of common understanding of the application domain. This stresses the important roles of developing a domain ontology, which can help establish shared understanding in a DSS domain (Corcho et al. 2003; Miah et al. 2014). Third and finally, this method should provide detailed guidelines for the DSS development, rather than general guidance or principles.

**Method Development.** Our method development found many insights from the development of the DTN, as reported by Pries-Heje and Baskerville (2008). In particular, the DTN provides conceptual guidance on how to connect alternative sources of knowledge, which is a topic relevant to our problem. However, we relaxed the requirement for using extant theory and instead explored the possibilities to connect different conceptual and ontological elements related to the DSS problem. The method does that through the SCOA framework, detailed in the following section.

## 4 SCOA

In this section we propose an IS artefact named SCOA, a method guiding DSS development in low-theoretical domains. SCOA defines four conceptual elements: scoping knowledge source, conceptual model, ontology, and DSS artefact. Although these elements have already been used in design science research, SCOA analyses and represents them in a heuristic way that can ground experience and expertise knowledge in the low-theoretical DSS domains. These elements are graphically represented in Figure 2. The sequent linkage between them is represented by full lines, i.e. the scoping knowledge source precedes the conceptual model, which precedes the ontology, and then leads to the artefact development. The dotted line represents the additional data flow between components. The next sections present further details about the components: goals, nature, how to process, and their linkage.

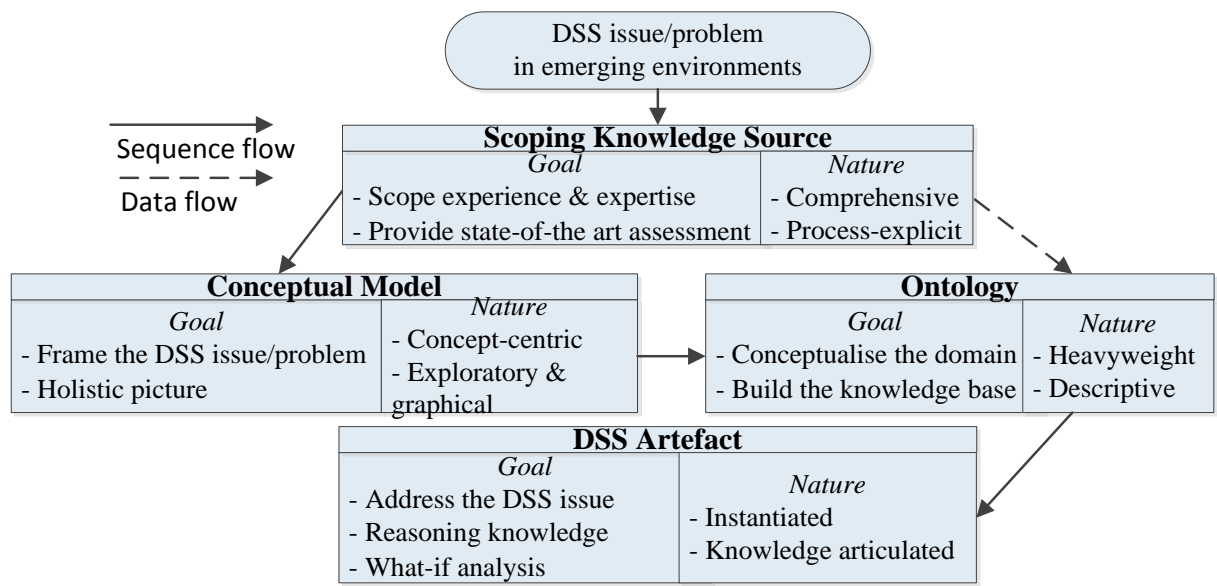

*Figure 2: The SCOA method*

### 4.1 Scoping Knowledge Source

Using the DSS issue/problem as the starting point, the scoping knowledge source comprehensively extracts and articulates existing experience and expertise related to the issue. Since the method targets low-maturity domains, where individual sources of knowledge predominate over established theory, reviewing these sources can help create a firm foundation for enhancing knowledge in IS research (Paré et al. 2015), and in particular design science research (Van Aken 2005). Another goal set by this component is to establish the relationships between the DSS issue and the state-of-the-art, which highlights the potential innovations brought by addressing the issue (Gregor and Hevner 2013; Peffers et al. 2007).

Regarding the specific nature of the component, we note that experience and expertise are highly reported in the scientific and practical literature. Thus, we suggest a literature review as the main part of this component. Although a narrative review could still be used, a scoping review is strongly suitable for the low-maturity emerging domains (Levac et al. 2010). More precisely, the 'scoping' classifier refers to a comprehensive sample strategy, which covers the vast range of experience and expertise in published and grey literature. Another benefit of scoping review is its explicit process of how the review is conducted (Levac et al. 2010; Paré et al. 2015), which increases the level of transparency and rigour of the to-be-built DSS artefact.

The IS field already has several guidelines to review the literature (Kitchenham 2007; Levy and Ellis 2006). In the SCOA method, we suggest following detailed accounts on how to conduct a scoping literature review proposed by Okoli and Schabram (2010) and Paré et al. (2015). These studies clearly represent the 'scoping' nature, and explicitly explain the review steps, including how to search, extract, and synthesise the related knowledge sources. Besides, we highlight three important points that should be considered in the SCOA method. First, to be comprehensive, as suggested by Paré et al. (2015), the searches should go beyond academic papers, for instance including practitioner and business surveys. The reason is that these reports may include relevant expert opinions, recommendations, and best practices, which are valuable sources of information in emergent application domains.

Second, when extracting data from the searched sources, it is noted that these data provide inputs for building the conceptual model and ontology. Thus, we suggest extracting data about the main concepts, relationships, processes, and rules related to the DSS issue. Of course, data extraction has to be related to the DSS issue, and thus be guided by the research questions (Rousseau et al. 2008). Third and finally, a good way to synthesise the extracted data is qualitative synthesis. Okoli and Schabram (2010) defines three types of synthesis: quantitative, qualitative, and a combination of them. Since our focus is on experience and expertise, qualitative synthesis that identifies patterns and thematic interpretations seems a more suitable approach. Furthermore, the pattern and theme analysis is also an emphasised characteristic of scoping literature reviews (Paré et al. 2015).

## 4.2 Conceptual Model

According to Cross (1982), design science involves the following activities: pattern-formation, synthesis, and modelling. While the above component formulates patterns and synthesises thematic interpretation, the conceptual model component starts the modelling activity. Using the data gathered with the scoping knowledge source, this component aims at developing a model that "simplified conceptualizations and representations of problems" (Hevner et al. 2004, p. 85). More precisely, it helps to grasp the main concepts in an application domain (Webster and Watson 2002), which in turn give researchers a holistic picture of the DSS issue (Cross 1982). From a design science perspective, a conceptual model built this way constitutes an IS artefact *per se* (Hevner et al. 2004). Thus, evaluation actions can be applied, using for example expert evaluation, which initially ensures the rigor of the to-be-built DSS artefact.

In SCOA, we view the conceptual model as a research framework that articulates the researcher's mind. Nunamaker and Chen (1990) suggest that a main activity of artefact development involves building a research framework that analyses the research issue in the related context. From this view, the framework is a substitute for a (little-t) theory, if we do not have it. Nevertheless, different levels of validity should be noted. While theories have a higher validity level, the framework checks the validity of what the researcher is doing. Using the framework, researchers are forced to make decisions on what information should be analysed and studied (Miles et al. 2014). Given that, although researchers can conceptualise the model in their minds, Miles et al. (2014) suggest a graphical presentation, which clearly specifies the concepts and relationships that hold the research issue. Furthermore, this graphical model supports the communication activities, especially where many researchers are involved.

The next concern is how to build the conceptual model. SCOA, based on Webster and Watson (2002), draws main concepts, factors, and processes from the scoping knowledge source to build and generalise a conceptual model. According to Webster and Watson (2002), extracted concepts may be classified according to the variance and process perspectives, but fortunately "[researchers] may draw from both variance and process research to develop conceptual models" (p. xix). Among the extracted concepts, the conceptual model should focus on the dominant ones to keep the model both focused and representative, which is aligned with the 'wisdom of the researchers' (Thuan et al. 2014). Thus, we suggest using the following table to synthesise the individual concepts. In this table, each concept is specified through its name, definition, references, and total number of sources addressing the concept. The number of supporting sources can be used to identify the dominant concepts related to the DSS issue. Based on Table 1, researchers should then ask for the links between the main concepts (Levy and Ellis 2006), which helps uncover the main conceptual relationships in the domain.

| Concepts (variance/process) | Description/Definition | References | Total of supporting sources |
|---|---|---|---|
| x | x-definition | x-references | x-total sources |

*Table 1: Structure for summarising main concepts related to the design problem*

After identifying the concepts and relationships, the next step is to graphically represent them in a conceptual model. We note that there could be different representations of the same group of identified concepts and relationships, depending on the adopted level of abstraction (Wieringa 2009). In the specific context of SCOA, we outline here a few recommendations for developing the conceptual model. First, the conceptual model should be abstract to give researchers an overall picture of the DSS issue. Otherwise, if a low level of abstraction is adopted, the model may become too complex, which is undesirable for a conceptual model (Jonker and Pennink 2009). Second, if some concepts and relationships are supported by many knowledge sources (identified via column 4 in Table 1), they should play a central role and thus be captured in the conceptual model. Finally, there may be more than one relationship between any two concepts. In these cases, we suggest modelling the one that emerges as predominant from the knowledge source. Other types of relationships may be specified in the next stage (e.g. in the ontology).

### 4.3   Ontology

The third distinctive component of the SCOA method is the ontology component, which is critical to consolidate knowledge in the emergent domain. Several studies have suggested that ontologies can improve knowledge structures and understanding in the application domains (Miah et al. 2014; Ostrowski et al. 2014). Furthermore, building an ontology in our method adds to the conceptual model in two important ways. First, it provides descriptive conceptualisation of the domain by explicitly articulating the concepts and relationships (Miah et al. 2007). As a result, the ontology can help resolve conflicts in the emerging domain by identifying whether several 'things' are represented by the same constructs (Shanks et al. 2003). Second, the ontology provides structured means for modelling the DSS knowledge base, as suggested by Miah and von Hellens (2014), who used ontologies for "structuring and representing problem specific knowledge [of] decision-making realities" (p. 261).

By and large, ontologies can be classified into lightweight ones, which only capture concepts and relationships in the domain, and heavyweight ones, which add rules and axioms that constrain ontological elements (Corcho et al. 2003; Valaski et al. 2012). In the SCOA method, we suggest building a heavyweight ontology, which adds reasoning prescriptive knowledge to the DSS artefact. From a design science perspective, the reasoning knowledge allows comparing among alternatives for given (sub) issues (Pries-Heje and Baskerville 2008). From a decision support perspective, the reasoning knowledge formulates decision parameters and prescriptive rules, and thus contributes to making informed decisions (Miah et al. 2014).

The ontology literature has proposed several methods for building ontologies (Corcho et al. 2003; Pinto and Martins 2004). By reviewing these methods, we suggest two steps that are popularly used and aligned with the SCOA method: ontology capture (Uschold and King 1995) and knowledge organisation (Küçük and Arslan 2014; López et al. 2004). The first step analyses extracted data for ontological elements: concepts, relationships, and axioms (data flow line in Figure 2). The analysis is similar to the analysis in the conceptual model but at a more detailed level. Additionally, software tools identifying ontological elements from text, like OntoGen (Fortuna et al. 2007), can also be used to analyse the extracted data. As a result, Table 1 is transformed into a glossary table capturing the ontological elements, as suggested by López et al. (2004).

The knowledge organisation step then synthesises the ontological elements and structurally arranges them into the ontology. In immature application domains, conflicted findings on these elements may be found. To address these conflicts, the synthesis follows the 'wisdom of the researchers' strategy, which states that the collective of researchers is smarter than the few (Thuan et al. 2014). Thus, the ontological elements supported by many sources should be included in the ontology, while the others should be eliminated. A similar strategy has already been applied to successfully develop ontologies in other areas (e.g. Osterwalder 2004). These ontological elements then need to be organised in a manageable way. Some overall structures that can be considered for this organisation are hierarchy structure, network structure, process structure, and layer structure. We note that this is an iterative process where we may extend, clean up, and update the ontology several times.

### 4.4   DSS Artefact

The final component of SCOA considers the artefact development, which should propose the artefactual solution addressing the DSS issue. In congruence with instantiation artefacts (Hevner et al. 2004), the developed artefact should be concrete and thus different from the conceptual models and ontologies. Such artefacts may be realised in a variety of forms, ranging from a simple spreadsheet to a much more complex decision tool (Thuan et al. 2015c). Within SCOA, the artefact should be built on the knowledge base provided by the ontology. More precisely, the artefact can use the ontological concepts for presenting its parameters, and the ontological relationships and axioms for guiding its reasoning. As a result, the instantiation artefact may automate a set of rules and constraints related with the decision process, and support the elaboration of different what-if scenarios.

If an artefact is developed in the form of software package, researchers need to choose an appropriate software development method that can be applied to the emerging fields. Given the dynamic nature of the emergent field, we suggest a rapid prototyping method, which addresses the dynamism by iteratively developing and revising a few software prototypes (Kordon 2002) and traversing the DSS issue design space (Lim et al. 2008). Furthermore, prototyping is appropriate for DSS development, as suggested by Miah et al. (2009) regarding the development of an expert system supporting rural business operators, and Antunes et al. (2014) regarding the development of a decision tool supporting geo-collaboration.

# 5 Evaluation

## 5.1 Demonstration

Following Peffers et al. (2007), we consider the assessment of the artefact as a composition of two activities: demonstration and evaluation. In this section, we demonstrate the proposed method by retroactively exercising it with an already published research project (Osterwalder 2004). The exercise shows how the project outcomes are aligned with the SCOA's components, which indicates the fitness between the SCOA method and the research process.

More precisely, the project focuses on business modelling, which was an emerging research area at that time and thus had no overarching theory (Osterwalder and Pigneur 2002). Similar to the SCOA method, the study conducted a review of the business model literature and used it as the main sources for developing a conceptual model and an ontology. In the literature review, the authors did not detail the review procedure, which, by default, would fall into the category of narrative literature review. Using the outcome of the review, these authors elaborated a conceptual model, which was named 'the 4 pillars of the business model ontology' (Osterwalder and Pigneur 2002). We believe that this model serves the exploratory purpose, as it framed the main concepts in the domain. Then, Osterwalder (2004) developed an ontology detailing the four pillars of the conceptual model, which was also influenced by the literature review (the data flow of the SCOA method). In particular, Osterwalder (2004) elected nine concepts that were "mentioned by at least two authors [references]" (p. 11) as the main concepts of the proposed ontology. Regarding artefact instantiation, a set of prototypes and an e-business model tool were developed (Osterwalder 2004; Osterwalder et al. 2002), which were based on the developed ontology.

The process adopted in this case is highly similar to the SCOA method, from reviewing the knowledge source, the conceptual model, ontology, and DSS artefacts. Both the conceptual model and ontology are based on data extracted from the review, while they together support the artefact development. Furthermore, the idea of using concepts suggested by two references is what we referred as the 'wisdom of the researchers'. Besides the major similarity, some minor differences between the SCOA method and the case can still be found. For instance, the case adopts a narrative review, whereas SCOA suggests scoping knowledge source, which is more rigorous (Okoli and Schabram 2010) and thus more suitable for design science.

## 5.2 Evaluation

Besides demonstration, Peffers et al. (2007) also suggest evaluating a design artefact in terms of how well it meets the design objectives, which can be achieved through actual use. In the current study, the SCOA method was evaluated by actually applying it to a real-world research project that aimed at developing a decision tool supporting the establishment of business process crowdsourcing (BPC) (Thuan et al. 2014; Thuan et al. 2015a; Thuan et al. 2015c). Since crowdsourcing is a low-theory area (Zhao and Zhu 2014), the SCOA method was chosen to guide the project. The project started by scoping 238 papers related to BPC, and analysed them for concepts and hierarchy relationships. The project applied the 'wisdom of the researchers' to choose main concepts and relationships, e.g. concepts suggested by at least 10 papers, which were used to develop a conceptual model (Thuan et al. 2014). The project then constructed an ontology of BPC by following the two-step process discussed in section 4.3. As a result, the ontology was structured as a four-layer framework: BPC components, their processes, data used in these processes, and data attributes. Furthermore, the ontology also included decision-making relationships and business rules, which turns it into a heavyweight ontology with the addition of reasoning knowledge (Thuan et al. 2015b). Based on the ontological knowledge base, the project developed a decision tool articulating the reasoning from the ontology, which is currently being tested (Thuan et al. 2015c).

Reflecting on this application, we now discuss the SCOA method in terms of the three pre-defined objectives: enable grounding of scattered knowledge, support shared understanding, and explicitly guide research activities. First, SCOA can ground knowledge from individual sources in the domain, which can be clearly seen via the scoping literature review (Thuan et al. 2014). Second, the method helps to achieve shared understanding. Indeed, the conceptual model and ontology include concepts, hierarchical relationships, decision making relationships, and business rules (Thuan et al. 2014; Thuan et al. 2015b), providing the necessary means to understand the BPC domain (Corcho et al. 2003). Another aspect of shared understanding includes the resolution of conflicts in the literature, which was handled by applying the 'wisdom of researchers'. Third, the SCOA method provides detailed steps to perform design science studies. This can be seen via its detailed description in section 4, and the fact that these steps were already adopted and worked well in the crowdsourcing project. Given this discussion and that several parts of the project have been published, we suggest the SCOA method has met its design objectives.

To sum up, Table 2 summarises the two aforementioned cases using the SCOA language.

| The business model | The decision tool of BPC |
|---|---|
| *Narrative review:* identify the research gap and explore the business model literature | *Scoping review:* identify and synthesise main components of BPC |
| *Explorative model:* constitute the essential business-model concepts | *Process model:* construct a process of BPC |
| *Formal heavyweight ontology:* provide a foundation for understanding and measuring business models | *Heavyweight ontology:* structure domain knowledge and capture business rules |
| *Prototyping installation:* verify the ontology and provide business modelling tool | *Knowledge articulated artefact:* design a decision tool supporting BPC establishment |

*Table 2: Summary the demonstration and evaluation cases using the SCOA language*

# 6 Conclusion

There is a need for DSS research in emerging application domains where theory is lacking and expertise and experience knowledge is dominant (Hosack et al. 2012). Addressing this need, we developed a design science method articulating several components of the research process, with the intention to collect and organise the experience and expertise on a particular application domain. The proposed method suggests beginning with a scoping knowledge source for extracting scattered, often-conflicting knowledge related to a design problem. Based on the extracted data, the method recommends building a conceptual model and then an ontology, for structuring the domain knowledge (Miah et al. 2014). Finally, the method brings the outcomes from these stages into DSS artefact development.

Following Peffers et al. (2007), the method is demonstrated through a published case (Osterwalder 2004) and evaluated by applying it to a crowdsourcing research project (Thuan et al. 2014; Thuan et al. 2015a; Thuan et al. 2015c). In the demonstration, we do not claim that the authors of the case would recognise that what SCOA offers are crucial to their success. Rather, we want to show that the components of SCOA have already been used and successfully addressed design problems in emerging domains. In the evaluation, we represented how the different components of SCOA can be actually used to conduct DSS research. This informs and justifies the sequent order proposed by the method, and thus initially confirms that the artefact can work in a practical real-world environment. We note that using illustrative cases is one of the most popular ways to demonstrate and evaluate design artefacts, especially design method artefacts (Peffers et al. 2012).

From a design science perspective, our study highlights the need for methods that can be applied to application domains with both low established theories and high scattered knowledge (quadrant 4 in Figure 1). We then propose SCOA, which is a first effort to develop a design science method addressing these domains. Using the method, researchers can build innovative artefacts, which in turn may be used as a research tool to understand the domain. Furthermore, we extend the work by Pries-Heje and Baskerville (2008), connecting a variety of knowledge sources (e.g. academic papers, expertise opinions, recommendations, experience, and best practice) to form a rigorous knowledge base for design science research.

We identify two main limitations in the current work. First, when considering SCOA within the context of the design cycle proposed by Hevner and Chatterjee (2010) that comprises both build and evaluation, the method is mainly located in the build activity. Second, we used two cases to demonstrate and evaluate the method, and the argument on generalizability of the method is certainly a valid one. Future research could address these limitations and extend the boundaries of the current work in several ways. Researchers may suggest combining SCOA with additional evaluation activity. In this case, the work on strategies for evaluation methods by Pries-Heje et al. (2008) can be seen as a starting point. Another interesting direction is to apply and validate SCOA in a variety of DSS application domains where expertise and experience are predominant, which may help increasing generalizability of the method. Finally, given that DSS research may be conducted in areas where theories may not be established yet, we strongly call for design methods that guide artefact development in these areas.

## Copyright